\documentclass[preprint,aps,showpacs]{revtex4}

\begin{document}
\draft

\title{Transmission Coefficient of Ballistic Quantum Dot}
\author{Lev G. Mourokh}
\address{Department of Physics and Engineering Physics, \\
Stevens Institute of Technology, Hoboken, NJ 07030 }
\author{Jonathan P. Bird}
\address{Department of Electrical Engineering \\
The State University of New York at Buffalo \\ Buffalo, NY
14260-1900 }
\date{\today}

\begin{abstract}
{We analyze the electron transmission through a ballistic quantum dot which is connected to two quantum point contacts. We demonstrate that the transmission and reflection of this structure is determined by the transfer time of the quantum-dot region which, in turn, is dependent on the applied electric field. The electric-field dependence of the transmission coefficient exhibits clearly pronounced peaks, and regions with negative differential transmission, and we discuss the conditions that should be satisfied in order to observe these effects in experiment.}
\end{abstract}

\pacs{73.63.Kv; 73.23.Ad; 85.35.Be }

\maketitle

Semiconductor nanostructures provide an excellent tool for theoretical and experimental investigations of electron dynamics, in the unique regime where the size of the structure and the number of particles are small enough to give rise to deviations from usual classical descriptions. In recent years, numerous publications have appeared, devoted to these so-called mesoscopic structures (see Refs. \cite{1,2,3,4}, and references therein). Moreover, the split-gate technique \cite{5} makes it possible to prepare structures of given shape. However, the majority of these non-classical phenomena have been examined for situations close to equilibrium, i.e. for the linear response regime, with nonequilibrium transport properties and their possible device applications remaining to be addressed. In this work we examine the electron transport through a ballistic quantum dot having length $L$ in $y$-direction and soft wall confinement potential with confinement frequency $\omega_0$ in $x$-direction (Figure 1). Initially (at time $t=0$), electrons are injected from the left point contact (region I) having confinement frequency $\omega_1$ larger than $\omega_0$. Subject to an electric field with strength $E$, electrons move through quantum dot (region II) and at time $t_T$ they reach the right point contact (region III). Our goal is to determine the transmission coefficient of this structure and its dependence on the applied electric field. After being injected from the left point contact, the electron wave function starts to spread out \cite{6}, but the confinement potential prevents this spreading and leads to oscillations in the width of the electron wave function. The relative parts of electron flow being reflected by the boundary between regions II and III or being transferred into region III depend on the phase of these oscillations and, consequently, on the transfer time, $t_T$, which is, in turn, is dependent on the applied electric field.

The electron wave functions in regions I and III are given by
\begin{equation}
\Psi_I = \psi (x,y,t=0) = {1\over \sqrt{2\pi}}e^{ik_1y}\left( {m\omega_1\over\pi\hbar}\right)^{1/4}\exp \left( -{m\omega_1\over 2\hbar}x^2\right)
\end{equation}
and
\begin{equation}
\Psi_{III} = \psi (x,y,t=t_T+0) = {1\over \sqrt{2\pi}}e^{ik_3y}\left( {m\omega_1\over\pi\hbar}\right)^{1/4}\exp \left( -{m\omega_1\over 2\hbar}x^2\right) ,
\end{equation}
respectively. Here, $\omega_1$ is the frequency of the electron confinement potential of the point contacts. The electron concentration in the point contacts is assumed to be low enough that only the first subband is occupied. In addition, we suppose that $\omega_1 \gg \omega_0$, so we can neglect the reflection on the boundary between regions I and II.

The time evolution of electron wave function in the region II is described by the following equation \cite{7}:
\begin{eqnarray}
\Psi_{II} = \psi (x,y,t) &=& \int dx_0dy_0K(x,y,t;x_0,y_0,0)\psi (x_0,y_0,0) \nonumber \\ &=&
\int dx_0K_{\omega}(x,t;x_0,0)\psi (x_0,0) \int dy_0K_E(y,t;y_0,0)\psi (y_0,0),
\end{eqnarray}
where $ K_{\omega}(x,t;x_0,0)$ is the Green's function of harmonic oscillator given by
\begin{equation}
K_{\omega}(x,t;x_0,0)=\left( {m\omega_0\over 2\pi i\hbar \sin\omega_0t}\right)^{1/2}\exp \left\{ {im\omega_0\over 2\hbar \sin\omega_0t} [(x^2+x_0^2)\cos\omega_0t-2x_0x]\right\}
\end{equation}
and $ K_E(y,t;y_0,0)$ is the Green's function of an electron subject to the electric field:
\begin{equation}
K_E(y,t;y_0,0) = \left( {m\over 2\pi i\hbar t}\right)^{1/2}\exp \left\{ {i\over \hbar} \left[ {m(y-y_0)^2\over 2t}+{1\over 2}eEt(y+y_0)-{eEt^3\over 24m}\right]\right\} .
\end{equation}

Electrons injected into the quantum dot from the left point contact are transmitted to the right point contact only partially due to reflection on the boundary between regions II and III. Consequently, the current density produced in the region III by electrons injected from region I is proportional to the transmission coefficient of this boundary squared. In turn, this transmission coefficient can be defined as an overlap integral of wave functions in regions II and III. As a result, we obtain
\begin{equation}
j_{out} = {e\hbar\over 2m}k_3 \vert\Psi_{III}\vert^2 \vert\langle\Psi_{III}\vert\psi (x,y,t_T)\rangle\vert^2.
\end{equation}
This overlap integral squared can be obtained from Eqs. (2-5) as
\begin{eqnarray}
\vert\langle\Psi_{III}\vert\psi (x,y,t_T)\rangle\vert^2 &=& T_ET_{\omega} = \delta \left( k_3-k_1-{eEt_T\over\hbar}\right) \nonumber \\
&&\times \left( \cos^2\omega_0t_T + {\sin^2\omega_0t_T\over 4\omega_1^2\omega_0^2}\left( {\omega_1^4\sin^2\omega_0t_T + \omega_0^2\omega_1^2 + \omega_0^4\cos^2\omega_0t_T\over \omega_1^2\sin^2\omega_0t_T + \omega_0^2\cos^2\omega_0t_T}\right)^2\right)^{-1/2} .
\end{eqnarray}
The transfer time $t_T$ is given by
\begin{equation}
t_T = {\hbar k_1\over eE} \left(\sqrt{1 + {2meEL\over\hbar^2k_1^2}} - 1\right) .
\end{equation}
It is evident from Eq.(6) that for the transfer time $t_T=\pi
n/\omega_0$, where $n$ is integer, the overlap integral is unity
for appropriate wave numbers $k_1$ and $k_3$, whereas for the
transfer time $t_T=\pi (2n+1)/2\omega_0$, this integral is about
$2\omega_0/\omega_1 \ll 1$. The transfer time depends on $E$
leading to the nonlinear dependence of the outcoming current on
the applied electric field strength. It should be noted that
electrons reflected on the boundary between regions II and III can
also be reflected on the boundary between regions II and I, and so
approach the region III again. However, the transverse shape of
their wave function would be the same as at the first approach
preventing them to entering region III and contributing to the
current.

To determine the transmission coefficient of the whole structure and its dependence on the applied electric field strength, we have to calculate the ratio of incoming and outcoming currents averaged over all possible values of wavevector $k_1$ (Fermi distribution):
\begin{equation}
T(E) = {\overline{j_{III}}\over \overline{j_I}} = {\int{dk_1
{k_1\sqrt{1+{2meEL\over\hbar^2k_1^2}}T_{\omega}\over \exp\left\{
(\hbar^2k_1^2/2m-E_f)/k_BT\right\} + 1}}\over\int{dk_1 {k_1\over
\exp\left\{ (\hbar^2k_1^2/2m-E_f)/k_BT\right\} + 1}}},
\end{equation}
where $E_f$ is the Fermi energy and $k_B$ is the Boltzmann
constant. In the absence of electric field the transmission
coefficient becomes oscillatory function of the Fermi energy as
shown in Figure 2. We have employed the following set of
parameters: $T=4K; \omega_0= 3\cdot 10^{12} s^{-1};
\omega_1=5\cdot 10^{15} s^{-1};$ and $L=0.4\mu m$. The field
dependence of this transmission coefficient is presented in Figure
3 for $E_f=10meV$. One can see that the transmission coefficient
has peaks at values of electric field associated with the relation
\begin{equation}
\omega_0\sqrt{{2mL\over eE}} = \pi n,
\end{equation}
where $n$ is integer. It is evident from this figure that there are regions of negative differential transmission and this structure can be employed as an active element in modern electronic devices.

Finally, we consider the conditions necessary to observe the
calculated effects in experiment. Since these effects are a
consequence of the coherent interference between electron partial
waves, which undergo multiple scattering from the confining
boundaries of the dot, it is necessary that electron
phase-coherence should be preserved over time scales $t \gg t_T$.
In recent years, a number of groups have investigated the factors
that limit phase coherence in open quantum dots
\cite{8,9,10,11,12,13}. Estimates for the phase-breaking time
($\tau_{\phi}$) of order $10^{-11} s$ have been found at 4 $K$,
which should be compared with a transit time of order $10^{-12} s$
for a typical GaAs dot with submicron dimensions. Application of a
non-zero electric field across the quantum dots will cause a
decrease in the electron phase-breaking time \cite{13}, but if the
field dependence of $\tau_{\phi}$ is sufficiently weak then it
should still be realistic to expect that we can indeed observe the
predicted non-linear effects. In fact, for parameters chosen in
the present paper, the peaks with $n=1,2,3$ should be observable.
These calculations have been performed for quite large quantum
dots with low electron density, so we expect that Coulomb
blockade, or other charging effects, should not be important.

In summary, we have examined the transmission coefficient of the
ballistic quantum dot placed between two point contacts. We have
shown that in the presence of soft harmonic confinement potential
the width of electron wave function (in the direction
perpendicular to the electron flow) is oscillating function of
time. Accordingly, the transmission and reflection of electrons
are determined by the transfer time which, in turn, depends on the
applied electric field. As a result, the field dependence of the
transmission coefficient has very pronounced peaks and regions
having negative differential transmission which can be used in the
electronic devices.

The authors would like to thank Anatoly Smirnov for valuable
discussions. L. G. M. also gratefully acknowledges support from
the Department of Defense, Grant No DAAD 19-01-1-0592.

\newpage
\begin{center}
Figure Captions
\end{center}

Figure 1. Schematic of the ballistic quantum-dot system placed
between two quantum point contacts.

Figure 2. The transmission coefficient of the ballistic quantum
dot as a function of the Fermi energy for zero applied electric
field.

Figure 3. The transmission coefficient of the ballistic quantum
dot as a function of an applied electric field.

\end{document}